# Joint Active and Passive Beamforming for RIS-aided MIMO Communications with Low-Resolution Phase Shifts

Nuno Souto, *Senior Member, IEEE*

*Abstract*—Practical hardware limitations often impose a reduced number of available phase shifts at the elements of a reconfigurable intelligent surface (RIS). Most works often assume continuous phase-shits at the RIS elements for the transmit and passive beamforming optimization, which can lead to substantial performance loss. Therefore, to harvest the gains of RIS-assisted multi-stream multiple-input multiple-output (MIMO) communications under realistic phase shifts, this letter proposes a problem formulation for the maximization of the achievable rate over the transmit precoder and RIS elements, which avoids an explicit discrete constraint while still incorporating its effect. To efficiently tackle the resulting problem when considering large arrays and RIS panels, an iterative algorithm is derived which comprises a sequence of simple projections. Simulation results demonstrate that the proposed design can be very effective, especially with low-resolution phase-shifts.

## I. Introduction

With the advent of demanding applications such as virtual/augmented reality, holographic projection, autonomous driving and tactile Internet, it is already clear that full support will only be possible in the future sixth generation (6G) of wireless systems [1]. One of the key technologies that is expected to help address the challenges in 6G networks corresponds to reconfigurable intelligent surfaces (RISs) [2]. RISs are artificial planar surfaces comprising many reconfigurable elements that can be programed to manipulate incident electromagnetic waves at a very high spatial resolution. This allows to improve overall system performance, for example by reducing interference or extending signal coverage which is critical when operating in the higher frequency bands envisioned for 6G (mmWave and THz) [3]. A RIS is thus an enabler for the realization of smart wireless environments where the wireless channel can also be optimized alongside the transmitter and receiver [4].

Due to its appealing features, RIS has attracted extensive research work in recent years, as summarized in [5]. In particular, several works have focused on the design of both the transmit and passive beamforming [6]-[8]. While these works assumed continuous phase shifts at the RIS elements, realizing hardware that can support continuous tuning of the phases of the individual RIS elements can be prohibitively expensive and unfeasible, especially when the number of elements is large [9]. Therefore, practical RISs often operate with a reduced number of phase shifts [10]. The simplest approach to cope with finite resolution phase shifts is to directly quantize the solutions obtained with a continuous based algorithm to the nearest discrete values, as adopted in the projected gradient method (PGM) based approach in [8]. While simple to implement, this rounding method can result in significant performance degradation, especially with low-resolution phase shifts [11]. Therefore, some recent works have considered the discrete-valued phase shifts constraint directly in the optimization. In [10], the authors considered a multiple-input single-output (MISO) scenario and represented the joint optimization problem as a linear program with binary variables. The optimal solution is then found using the branch-and-bound method which has exponential computational complexity. The authors in [12] presented a low complexity algorithm that computes the optimal beamforming vector for any discrete phase resolution. However only single-antenna transmitters and receivers were considered. In [13], particle swarm optimization was applied as a heuristic method for rate maximization in scenarios with multiple single-antenna users, considering both continuous and discrete phase-shifts. A different approach relies on the adoption of a codebook-based framework [14]. Along this line [15] proposed a channel estimation and passive beamforming design where the reflection coefficients are selected from a randomly generated codebook having discrete phase shifts.

Instead of restricting the transmission to single stream and to a single antenna receiver, in this letter we study a multi-stream MIMO system where a RIS with a limited number of phase shifts helps to establish the communication between a transmitter and a receiver. Specifically, the main contributions of the paper can be summarized as follows:
- We target the joint maximization of the achievable rate over the transmit precoder and RIS under the assumption of discrete-valued phase shifts. To address the difficult mixed-integer non-linear program (MINLP) associated with the intended optimization, we reformulate the problem using the convex hull of the discrete phase shifts which allows us to remove the explicit discrete constraint

This work is funded by FCT/MCTES through national funds and when applicable co-funded by EU funds under the project UIDB/50008/2020.

N. Souto is with the ISCTE-University Institute of Lisbon and Instituto de Telecomunicações, 1649-026 Lisboa, Portugal (e-mail: nuno.souto@lx.it.pt).

while still incorporating its effect.
- To efficiently handle large problem settings, we apply the cyclic block proximal gradient (CBPG) method with an additional extrapolation step in the update of the individual variables. Combined with the adopted problem formulation this strategy allows us to derive a novel iterative algorithm comprising a sequence of simple projections.
- Through numerical simulations, we evaluate the performance of the proposed approach and show that it can be very effective, especially when low-resolution phase shifters are assumed at the RIS.

*Notation:* Matrices and vectors are represented by uppercase and lowercase boldface letters, respectively. $(\cdot)^T$ and $(\cdot)^H$ denote the transpose and conjugate transpose, $\otimes$ is the Kronecker product, $\mathbb{E}\{\cdot\}$ is the expected value, $\|\cdot\|_2$ is the 2-norm, $\|\cdot\|_F$ is the Frobenius norm, $\text{vec}(\cdot)$ is the vectorization operator which is a vector obtained by stacking the columns of the argument matrix, $\text{diag}(\cdot)$ transforms a vector into a diagonal matrix or represents the diagonal of a matrix, $\mathbf{I}_n$ is the $n \times n$ identity matrix, $\vec{\mathbf{1}}$ is the all-ones vector and $\mathcal{I}_\mathcal{D}(\mathbf{v})$ is the indicator function which returns 0 if $\mathbf{v} \in \mathcal{D}$ and $+\infty$ otherwise.

## II. SYSTEM MODEL

Let us consider the MIMO communication system illustrated in Fig. 1, where a transmitter equipped with $N_{tx}$ antennas sends $N_s$ simultaneous data streams to a receiver with $N_{rx}$ antennas, with the aid of a RIS panel comprising $N_{ris}$ elements. Representing the transmitted signal as $\mathbf{s} = [s_1 ... s_{N_s}]^T$, where $s_i \in \mathbb{C}$ is an amplitude and phase modulated symbol with $\mathbb{E}\{\|\mathbf{s}\|_2^2\} = N_s$, then the received signal can be expressed as

$$\mathbf{r} = \sqrt{\rho}\mathbf{H}\mathbf{F}\mathbf{s} + \mathbf{n}, \quad (1)$$

where $\sqrt{\rho}$ is the power per stream, $\mathbf{H} \in \mathbb{C}^{N_{rx} \times N_{tx}}$ is the channel matrix and $\mathbf{F} \in \mathbb{C}^{N_{tx} \times N_s}$ is the precoder matrix. Vector $\mathbf{n} \in \mathbb{C}^{N_{rx} \times 1}$ contains random noise samples distributed according to $\mathcal{CN}(\mathbf{0}, \sigma_n^2 \mathbf{I}_{N_{rx}})$. Matrix $\mathbf{H}$ can be written as

$$\mathbf{H} = \mathbf{H}_{S,D} + \mathbf{H}_{R,D} \boldsymbol{\Phi} \mathbf{H}_{S,R}, \quad (2)$$

where $\mathbf{H}_{S,D} \in \mathbb{C}^{N_{rx} \times N_{tx}}$ represents the direct channel between the transmitter and receiver, $\mathbf{H}_{S,R} \in \mathbb{C}^{N_{ris} \times N_{tx}}$ is the channel between the transmitter and the RIS and $\mathbf{H}_{R,D} \in \mathbb{C}^{N_{rx} \times N_{ris}}$ is the channel between the RIS and the receiver. $\boldsymbol{\Phi} \in \mathbb{C}^{N_{ris} \times N_{ris}}$ is the phase shift matrix of the RIS which has a diagonal structure namely, $\boldsymbol{\Phi} = \text{diag}(\boldsymbol{\varphi})$ with $\boldsymbol{\varphi} = [\varphi_1, ..., \varphi_{N_{ris}}]^T$. Assuming discrete phase shifts with $N_b$ quantization bits then each element of the RIS matrix, $\varphi_i$ $(i=1,...,N_{ris})$, can take $M = 2^{N_b}$ possible phase shift values.

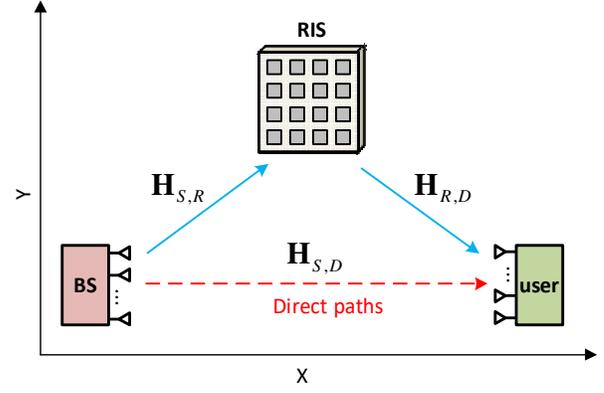

Fig. 1. System model layout considered for the multi-stream RIS-assisted MIMO communication system.

## III. PROBLEM FORMULATION AND ALGORITHM DESCRIPTION

### A. Achievable Rate Maximization with Discrete Constraints

Assuming Gaussian signaling and perfect channel knowledge, we can formulate the joint rate maximization problem as

$$\min_{\substack{\mathbf{F} \in \mathbb{C}^{N_{tx} \times N_s} \\ \boldsymbol{\varphi} \in \mathbb{C}^{N_{ris} \times 1}}} f(\mathbf{F}, \boldsymbol{\varphi}) = -\ln \det\left(\mathbf{I}_{N_s} + \frac{\rho}{\sigma_n^2}\mathbf{F}^H \mathbf{H}^H \mathbf{H} \mathbf{F}\right) \quad (3a)$$

$$\text{subject to } \|\mathbf{F}\|_F^2 \leq N_s \quad (3b)$$

$$\varphi_i \in \left\{a, a \cdot e^{j\frac{2\pi}{M}}, ..., a \cdot e^{j\frac{2\pi(M-1)}{M}}\right\}, \ i=1,...,N_{ris}. \quad (3c)$$

where $a$ denotes the amplitude of the reflection coefficients. The coupling between $\mathbf{F}$ and $\boldsymbol{\varphi}$ in (3a), combined with the discrete constraint (3c), results in a nonconvex MINLP whose global solution is difficult to find efficiently. To make the problem simpler to handle, we can relax (3c) using the convex hull of the discrete phase shifts. First, using $\theta_k = a \cdot e^{j 2\pi k / M}$ we rewrite (3c) as

$$\varphi_i = \sum_{k=0}^{M-1} t_{i,k} \theta_k, \ t_{i,k} \in \{0,1\}, \ \sum_{k=0}^{M-1} t_{i,k} = 1. \quad (4)$$

Relaxing the second condition to $t_{i,k} \geq 0$ we can rewrite (4) as

$$\varphi_i = \mathbf{t}_i^T \boldsymbol{\theta}, \ \mathbf{t}_i \geq 0, \ \mathbf{t}_i^T \vec{\mathbf{1}} = 1, \ i=1,...,N_{ris} \quad (5)$$

where $\boldsymbol{\theta} = [\theta_0, ..., \theta_{M-1}]^T$ and $\mathbf{t}_i = [t_{i,0}, ..., t_{i,M-1}]^T$. Applying this convex set constraint and defining $\mathbf{T} = [\mathbf{t}_1, ..., \mathbf{t}_{N_{ris}}]^T$ and $\mathbf{t} = \text{vec}(\mathbf{T}^T)$, we can reformulate problem (3) as

$$\min_{\substack{\mathbf{F} \in \mathbb{C}^{N_{tx} \times N_s} \\ \mathbf{T} \in \mathbb{R}^{N_{ris} \times M}}} f(\mathbf{F}, \mathbf{T}\boldsymbol{\theta}) + \mathcal{I}_\mathcal{C}(\mathbf{F}) + \mathcal{I}_\mathcal{D}(\mathbf{t}), \quad (6)$$

which uses sets $\mathcal{C} = \{\mathbf{F} \in \mathbb{C}^{N_{tx} \times N_s} : \|\mathbf{F}\|_F^2 \leq N_s\}$ and $\mathcal{D} = \{\mathbf{t} \in \mathbb{R}^{MN_{ris}} : \mathbf{t} \geq 0, \mathbf{t}_i^T \vec{\mathbf{1}} = 1\}$.

### B. Proposed Algorithm

To address problem (6) while coping with the typical large

**Algorithm 1:** DA-CBPG Algorithm

1: **Input:** $\mathbf{r}$, $\mathbf{H}$, $\alpha$, $Q$, $\mathbf{F}^{(0)}$, $\mathbf{t}^{(0)}$
2: $\mathbf{P}^{(0)} = \mathbf{F}^{(0)}$, $\mathbf{y}^{(0)} = \mathbf{t}^{(0)}$
3: **for** $q=1,\ldots,Q$ **do**
4: $\quad \mathbf{F}^{(q+1)} = \text{prox}_{\alpha \mathcal{I}_\mathcal{C}}\left(\mathbf{P}^{(q)} - \alpha \nabla_{\mathbf{F}^*} f\left(\mathbf{P}^{(q)}, \left(\mathbf{I}_{N_{ris}} \otimes \boldsymbol{\theta}^T\right)\mathbf{t}^{(q)}\right)\right)$
5: $\quad \mathbf{P}^{(q+1)} = \mathbf{F}^{(q+1)} + \frac{q}{q+3}\left(\mathbf{F}^{(q+1)} - \mathbf{F}^{(q)}\right)$
6: $\quad \mathbf{t}^{(q+1)} = \text{prox}_{\alpha \mathcal{I}_\mathcal{D}}\left(\mathbf{y}^{(q)} - \alpha \nabla_{\mathbf{t}} f\left(\mathbf{F}^{(q+1)}, \left(\mathbf{I}_{N_{ris}} \otimes \boldsymbol{\theta}^T\right)\mathbf{y}^{(q)}\right)\right)$
7: $\quad \mathbf{y}^{(q+1)} = \mathbf{t}^{(q+1)} + \frac{q}{q+3}\left(\mathbf{t}^{(q+1)} - \mathbf{t}^{(q)}\right)$
8: **end for**
9: **Output:** $\mathbf{F}^{(q+1)}$, $\boldsymbol{\varphi} = \left(\mathbf{I}_{N_{ris}} \otimes \boldsymbol{\theta}^T\right)\mathbf{t}^{(q+1)}$.

dimensions in RIS-aided MIMO systems, we apply the CBPG method [16] as it is a first order optimization method that allows a simplified computation of matrices $\mathbf{F}$ and $\mathbf{T}$ in turn, keeping the other variable fixed. Each update consists of a gradient-based step, followed by a proximal mapping, defined as

$$\text{prox}_{\alpha g}(\mathbf{a}) = \arg\min_{\hat{\mathbf{x}}}\left(g(\hat{\mathbf{x}}) + \frac{1}{2\alpha}\|\hat{\mathbf{x}} - \mathbf{a}\|_2^2\right). \quad (7)$$

for a function $g$. Instead of computing the gradient steps over the previous points, $\mathbf{F}^{(q)}$ and $\mathbf{t}^{(q)}$ ($q$ is the iteration number), we use extrapolated variables, $\mathbf{P}^{(q)}$ and $\mathbf{y}^{(q)}$, which are linear combinations of the previous two estimates, as these can improve the typical slow convergence of proximal gradient methods, as discussed in [17]. Based on remark (10.35) in [16], we set the extrapolation parameter as $q/(q+3)$. The overall discrete accelerated CBPG based (DA-CBPG) algorithm is summarized in Algorithm 1. $Q$ denotes the maximum number of iterations and $\alpha$ is the step size, which can be found through backtracking line search [18]. According to (7), the proximal mapping in step 4 reduces to the Euclidean projection over $\mathcal{C}$

$$\text{prox}_{\alpha \mathcal{I}_\mathcal{C}}(\mathbf{Z}) = \begin{cases} \mathbf{Z}, & \|\mathbf{Z}\|_F^2 \leq N_s \\ \frac{\sqrt{N_s}}{\|\mathbf{Z}\|_F}\mathbf{Z}, & \text{otherwise} \end{cases}, \quad (8)$$

defined for any matrix argument $\mathbf{Z}$. Step 6 can be expressed as $\text{prox}_{\alpha \mathcal{I}_\mathcal{D}}(\mathbf{x}) = \left[\Pi_\mathcal{S}(\mathbf{x}_1)^T \ldots \Pi_\mathcal{S}(\mathbf{x}_{N_{ris}})^T\right]^T$, where $\mathbf{x}$ and $\mathbf{x}_i$ represent generic vectors with the same sizes as $\mathbf{t}$ and $\mathbf{t}_i$ and $\Pi_\mathcal{S}(\cdot)$ is the projection over the probability simplex ([17], section 6.2), i.e.,

$$\Pi_\mathcal{S}(\mathbf{x}_i) = \left(\mathbf{x}_i - v_i\vec{\mathbf{1}}\right)_+, \quad (9)$$

where $v_i$ is the value that satisfies $\vec{\mathbf{1}}^T\left(\mathbf{x}_i - v_i\vec{\mathbf{1}}\right)_+ = 1$, with $(\cdot)_+$ defined as $(z)_+ = \max(z, 0)$. The algorithm in [19] can be used for computing (9) efficiently. The gradients required for steps 4 and 6 are computed according to the following lemma:

*Lemma 1:* Let $f(\mathbf{F}, \boldsymbol{\varphi})$ be defined as in (3a), where the channel is represented as (2) with $\boldsymbol{\varphi} = \left(\mathbf{I}_{N_{ris}} \otimes \boldsymbol{\theta}^T\right)\mathbf{t}$. Then, the gradients with respect to $\mathbf{F}^*$ and $\mathbf{t}$ can be obtained as

$$\nabla_{\mathbf{F}^*} f(\mathbf{F}, \boldsymbol{\varphi}) = -\frac{\rho}{P_n}\mathbf{H}^H\mathbf{HF}\left(\mathbf{I}_{N_s} + \frac{\rho}{\sigma_n^2}\mathbf{F}^H\mathbf{H}^H\mathbf{HF}\right)^{-1} \quad (10)$$

and

$$\nabla_{\mathbf{t}} f\left(\mathbf{F}, \left(\mathbf{I}_{N_{ris}} \otimes \boldsymbol{\theta}^T\right)\mathbf{t}\right) = -2\operatorname{Re}\left\{\operatorname{vec}\left(\boldsymbol{\theta}\operatorname{diag}^T(\mathbf{Z})\right)\right\}, \quad (11)$$

with

$$\mathbf{Z} = \mathbf{H}_{S,R}\mathbf{F}\left(\mathbf{I}_{N_s} + \frac{\rho}{\sigma_n^2}\mathbf{F}^H\mathbf{H}^H\mathbf{HF}\right)^{-1}\mathbf{F}^H\mathbf{H}^H\mathbf{H}_{R,D}. \quad (12)$$

*Proof:* Proof is shown in Appendix A.

### C. Computational Complexity

Apart from some small matrix inversions, the proposed DA-CBPG algorithm depends mostly on matrix and vector multiplications. It can be shown that the resulting computational complexity order is $\mathcal{O}\left(Q\left(N_{rx}N_{tx}N_s + N_{ris}M + N_b M + N_{rx}N_{tx}N_{ris}\right)\right)$ which means that it grows linearly with $M$ and $N_{ris}$. For comparison, an optimal algorithm which solves (3) through exhaustive search requires $\mathcal{O}\left(M^{N_{ris}}\left(N_{rx}^2 N_{tx} + N_{rx}N_{tx}N_{ris}\right)\right)$ operations, the quantized PGM algorithm from [8] has a complexity of $\mathcal{O}\left(Q_{PGM}\left(N_{rx}^3 + N_{rx}^2 N_{tx} + N_{rx}N_{tx}^2 + N_{rx}N_{tx}N_{ris} + N_{tx}^3\right)\right)$ whereas $\mathcal{O}\left(N_{rx}^2 N_{tx} + Q_{Succ}\left(N_{rx}N_{tx}^2 M + N_{rx}N_{tx}N_{ris}\right)\right)$ operations are required for a singular valued decomposition (SVD) based successive refinement algorithm extended directly from [10] (generalized to a multi-stream single-user MIMO transmission where the objective function is the achievable rate (3a)). Since the latter requires at least $Q_{Succ} = N_{ris}$ iterations to run over all RIS elements at least once, its complexity grows linearly with $M$ and quadratically with $N_{ris}$. PGM grows only linearly with $N_{ris}$ but has a cubic dependency on the transmitter and receiver array sizes.

### IV. NUMERICAL RESULTS

In this section we provide numerical results obtained using Monte Carlo simulations. We consider a multi-stream MIMO system operating at a frequency of 28 GHz, with a bandwidth of 800 MHz, with $\sigma_n^2 = 85$ dBm and where the RIS elements have a reflection amplitude of $a = 1$. We adopt a clustered geometric channel model [20] where the transmitter-RIS and RIS-receiver links have one line-of-sight (LOS) and $N_{ray}=10$ non line-of-sight (NLOS) components. The energy ratio between the LOS and NLOS components is set as $K_{Rice}=10$. In the case of the direct link, we assume that no LOS path exists due to surrounding obstructions. A path loss exponent of 1.90 is applied for the LOS propagation and 4.39 for the NLOS propagation [21]. Furthermore, square uniform planar arrays (UPAs) are used, with an inter-element spacing of $\lambda/2$, where $\lambda$ denotes the wavelength. The azimuth and elevation angles of departure and arrival are uniformly random distributed in $[-\pi,\pi)$ and $[-\pi/2,\pi/2)$, respectively. All the results are averaged over 1000 random channel realizations.

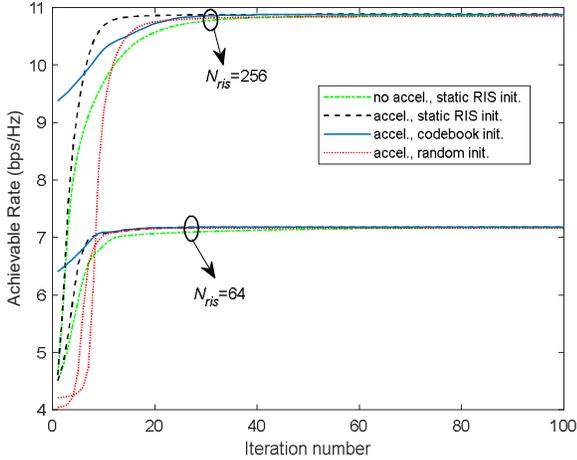

Fig. 2. Progress of the proposed DA-CBPG algorithm in terms of the achievable rate. Scenario with: $N_s$=2, $N_{tx}$=64, $N_b$=2, $N_{rx}$=16 and $P_{tx}$=30dBm.

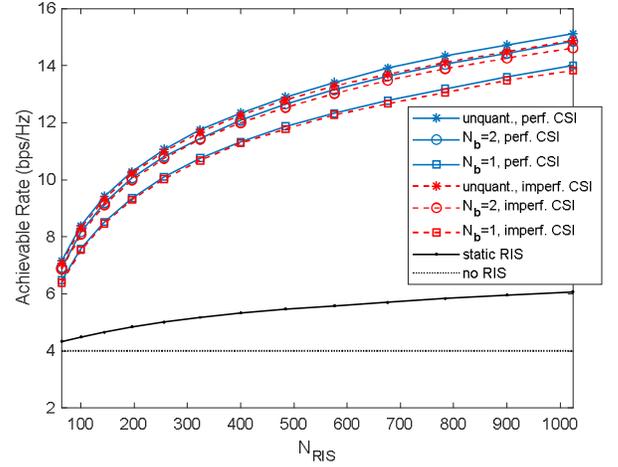

Fig. 3. Achievable rate versus number of RIS elements ($N_{ris}$) of DA-CBPG algorithm. Scenario with: $N_s$=2, $N_{tx}$=64, $N_{rx}$=16 and $P_{tx}$=30dBm.

First, we show the convergence of the proposed DA-CBPG algorithm in Fig. 2. The transmitter, receiver and RIS are located at (0,0) m, (40,0) m and (30,10) m, respectively (the third dimension is assumed to be the same for all). It can be observed that the accelerated algorithm takes 10-20 iterations to converge. In fact, when using the same initialization, namely a static RIS with all the elements set as $\varphi_i$=1, the use of the extrapolation step (accelerated version) allows the algorithm to converge to a solution in fewer iterations than the non-accelerated case. Furthermore, it can also be observed that different initializations for the RIS, such as the codebook initialization which adopts the Hadamard matrix (as in [22]) or a random initialization, do not have a noticeable impact on the final solution. Still, these may slightly change the number of required iterations for the algorithm to converge.

Considering the same scenario, Fig. 3 shows the achievable rate versus the number of RIS elements obtained using the proposed approach with $N_b$=1, $N_b$=2 and without quantization. The later corresponds to applying directly the CBPG method to problem (3) with constraint (3c) replaced by $|\varphi_i|=a$ (which results in a projection onto the set of vectors whose elements have modulus equal to $a$ as in the RIS-only accelerated proximal gradient (APG) algorithm from [6]). It can be seen that the use of a properly optimized RIS can provide substantial gains over a communication without RIS or with a static RIS. As expected, these gains become greater when adopting RISs with a larger number of elements. It is also visible that a performance degradation of 1-2 bps/Hz occurs when phase shifters with only one bit resolution are used. This loss reduces to under 0.5 bps/Hz when increasing the number of quantization bits to 2. Curves with imperfect channel state information (CSI) are also included in Fig. 3. To obtain these, it was assumed that each estimated channel matrix ($\mathbf{H}_{S,R}$, $\mathbf{H}_{R,D}$ and $\mathbf{H}_{S,D}$) is represented as the sum of the true channel and an estimation error matrix whose components are independent and identically distributed according to $\mathcal{CN}(0,\sigma^2)$, with $\sigma^2 = 0.25 \cdot \mathbb{E}\{|H_{i,j}|^2\}$. Even with a relatively large estimation error, the loss in the achievable rate stays under 0.5 bps/Hz.

Considering $N_b$=1, Fig. 4 plots the achievable rate versus transmit power for two different panel sizes. The receiver and RIS are located at (20,0) m and (18,2) m, and there is no direct link. In this figure, the proposed DA-CBPG algorithm is compared against the quantized PGM from [8], a Hadamard-based codebook scheme [10][22], an SVD-based successive refinement algorithm extended from [10] and the optimal solution. Due to the prohibitive high complexity of the latter approach, we only include the optimal curve for the smaller RIS. The best results are achieved by the methods that directly consider discretized phase shifts, namely DA-CBPG and the successive refinement algorithm. In the case of the smaller RIS, both approaches perform very close to the optimal solution, but DA-CBPG requires only $6\times10^5$ complex-valued multiplications whereas the latter requires $2\times10^6$. With the larger RIS, the DA-CBPG algorithm achieves the highest rates and has the lowest complexity with $6\times10^6$ operations, which compares against $24\times10^6$ of PGM and $35\times10^6$ of successive refinement.

## V. CONCLUSIONS

This paper addressed the problem of maximizing the achievable rate jointly over the transmit precoder and RIS elements, considering multi-stream MIMO communications with realistic discrete phase-shifts. As this optimization renders a difficult nonconvex MINLP, we proposed a different problem formulation, where the discrete-valued phase shifts constraint is approximated using the convex hull, upon which we could then apply the CBPG method combined with an extrapolation step. This allowed us to derive a novel iterative algorithm comprising a sequence of simple projections. Numerical results showed that the proposed approach can reduce the performance loss incurred with the use of low-resolution phase shifts at the RIS, achieving higher rates than other benchmarked schemes. While this work focused on point-to-point MIMO transmissions it should be possible to extend the approach to multiuser/multi-cell settings.

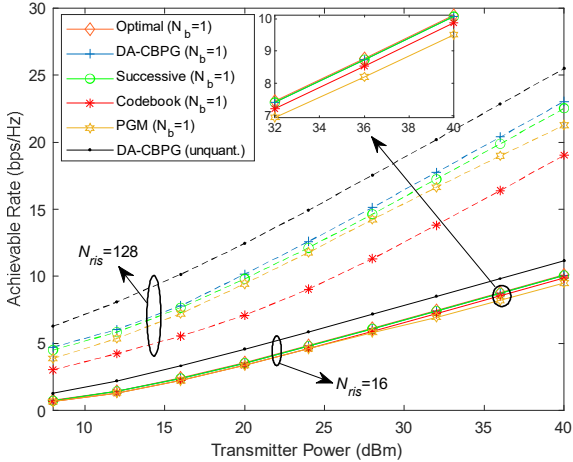

Fig. 4. Achievable rate versus transmitter power with different algorithms. Scenario with: $N_s=2$, $N_{tx}=64$, $N_{rx}=16$ and no direct link.

## APPENDIX A
### PROOF OF LEMMA 1

In the following we apply the procedure described in [23]. For obtaining the complex-valued gradient of $f(\mathbf{F}, \boldsymbol{\varphi})$ with respect to $\mathbf{F}^*$, we write the complex differential as

$$df = -\text{Tr}\left\{\frac{\rho}{P_n}\left(\mathbf{I}_{N_s} + \frac{\rho}{P_n}\mathbf{F}^H\mathbf{H}^H\mathbf{H}\mathbf{F}\right)^{-1} \times \left(\mathbf{F}^H\mathbf{H}^H\mathbf{H}d\mathbf{F} + d\mathbf{F}^H\mathbf{H}^H\mathbf{H}\mathbf{F}\right)\right\}. \quad (13)$$

Applying the relation $d(\ln\det(\mathbf{X})) = \text{Tr}\{\mathbf{X}^{-1}d\mathbf{X}\}$ ([23], Table 3.1) we can rewrite the differential as

$$df = -\text{Tr}\left\{\frac{\rho}{P_n}\left(\mathbf{I}_{N_s} + \frac{\rho}{P_n}\mathbf{F}^H\mathbf{H}^H\mathbf{H}\mathbf{F}\right)^{-1}\mathbf{F}^H\mathbf{H}^H\mathbf{H}d\mathbf{F} + \frac{\rho}{P_n}\left(\mathbf{H}^H\mathbf{H}\mathbf{F}\left(\mathbf{I}_{N_s} + \frac{\rho}{P_n}\mathbf{F}^H\mathbf{H}^H\mathbf{H}\mathbf{F}\right)^{-1}\right)^T d\mathbf{F}^*\right\}, \quad (14)$$

from which, using ([23], Table 3.2), allows us to directly obtain (10). Adopting a similar approach, we can write the differential of $f(\mathbf{F}, (\mathbf{I}_{N_{ris}} \otimes \boldsymbol{\theta}^T)\mathbf{t})$ with respect to $\mathbf{t}$ as

$$df = -\text{diag}^T(\mathbf{Z})d\boldsymbol{\varphi} - \text{diag}^T(\mathbf{Z}^H)d\boldsymbol{\varphi}^*, \quad (15)$$

where $\mathbf{Z}$ is defined as (12). Considering that $d\boldsymbol{\varphi} = (\mathbf{I}_{N_{ris}} \otimes \boldsymbol{\theta}^T)d\mathbf{t}$, we can write

$$df = -2\text{Re}\left\{\text{diag}^T(\mathbf{Z})(\mathbf{I}_{N_{ris}} \otimes \boldsymbol{\theta}^T)\right\}d\mathbf{t}, \quad (16)$$

which, results in the gradient expression in (11). ∎